\documentstyle[12pt,epsf]{article}

\font\rmtit=cmr12
\newcommand{\bce}{\begin{center}}
\newcommand{\ece}{\end{center}}
\newcommand{\beq}{\begin{equation}}
\newcommand{\eeq}{\end{equation}}

\newcommand{\gbg}{g_{\scriptscriptstyle{BG}}(r)}
\newcommand{\mref}[1]{(\ref{#1})}

\def\xb{{\bf x}}

\def\kb{{\bf k}}

\def\cG{{\cal G}}
\def\om{\omega}

\def\hN{{\hat N}}
\def\hrho{{\hat\rho}}

\def\eps{\varepsilon}

\begin{document}
\title{\hfill {\rmtit DFTT 15/95}\\
       \vspace{-0.3cm}\hfill {\rmtit February 1995}\\
\vspace*{1.cm}
{\bf Deconfinement transition \\in a one--dimensional model}}
\author{W.M. Alberico, M. Nardi and S. Quattrocolo \vspace*{0.3cm}\\
{\it Dipartimento di Fisica Teorica}\\
{\it dell'Universit\`a di Torino}\\
{\it and}\\
{\it INFN, Sezione di Torino}\\
{\it via P.Giuria 1, 10125 Torino, Italy}}
\maketitle

\begin{abstract}
\noindent
We present a model for quark matter with a density dependent quark--quark
(confining) potential,  which allows to describe a deconfinement phase
transition as the system evolves from a low density assembly of bound
structures to a high density free Fermi gas of quarks. A proper account of
the many--body correlations induced by the medium is crucial in order to
disentangle this behaviour, which does not uniquely stem from the
the naive density dependence of the interaction. We focus here on the
effects of finite (non--zero) temperatures on the dynamical behaviour of
the system. In particular we investigate the ground state energy per
particle and the pair correlation function, from which one can extract the
relevant information about the quarks being confined or not; the
temperature dependence of the transition density is also derived.
\end{abstract}

\section{Introduction}
``QCD inspired'' models for quark/nuclear matter have been proposed since
many years, mainly within the framework of non--relativistic constituent
quark models \cite{Hor1,Hor2,Hor3,abmp};
the latter have proven remarkably useful for
single--hadron spectroscopy, in spite of the lack of Lorentz invariance and
chiral symmetry. Relativistic models, based on effective chirally invariant
Lagrangians of NJL type or related to the so--called colour dielectric model,
have also provided important tools for the investigation of quark
matter \cite{Fried,Li,Chan,Pir,Pie};
however their many--body description mainly relies on the
mean field approach, which does not seem the most appropriate one to deal
with the problem of a quark--hadron phase transition.

The present work is based on a non--relativistic model for a one--dimensional
many quark system, whose key feature is a density dependent interaction
between quarks \cite{abmp}:
%
\beq V(x)={1\over 2}x^2\,e^{-c\rho |x|}\,,
\label{potenziale}
\eeq
$x\equiv x_1-x_2$ being the relative interquark distance, $\rho=N/L$ the
(uniform) density of the system with $N$ fermions in a length $L$, and $c$
a constant parameter. The potential \mref{potenziale}
resembles a strong confining force
in the limit of very low densities, where one expects quarks to be bound
into hadrons, while it becomes negligible at large densities, where the
quarks behave as a free Fermi gas (which should mimic the quark--gluon
plasma phase).

Internal degrees of freedom are neglected, but antisymmetry for the global
wavefunction of the system is required, as it is appropriate for an assembly
of fermions. We notice that while the many--body interaction potential of
\cite{Hor1,Hor2,Hor3}
 can only be dealt with variational Monte Carlo techniques, the
present two--body potential allows to describe the system within the
customary many--body schemes, which have been widely tested against nuclear
matter properties.

The philosophy underlying this model is to provide a tool for a
phenomenological but microscopic description of the phase transition from a
plasma of (weakly interacting) quarks and gluons to an assembly of colourless
quark clusters, namely hadrons. This transition should have occurred in the
early stages of the Universe, at rather large temperature and density of the
primordial gas of elementary constituents; as the system expanded and cooled
down, hadronization took place, giving rise to the ordinary matter of baryons
and mesons, eventually bound into nuclear systems.

Relativistic heavy ion collisions are presently investigated in an attempt of
reproducing the extreme density and temperature conditions at which the
quark--gluon plasma phase can occur. One can thus envisage that temperature
plays a crucial role in the description of the dynamical evolution of the
system. Indeed, according to lattice QCD calculations, confining forces
become weaker with increasing temperature, as it is suggested by the
following temperature dependence of the string tension \cite{Pis}:
%
\begin{equation}
\sigma(T)=\sigma_0\sqrt{1-\left({T\over T_C}\right)^2}\,\theta(T_C-T),
\label{tens}
\end{equation}
where $\sigma_0=\sigma(T=0)$ and the theta function implies that confinement
no longer survives above the critical temperature $T_C$. Obviously this
can only be taken as a {\it qualitative} indication of how the temperature
affects the quark--quark interaction potential; moreover various uncertainties
concern the precise determination of the critical temperature (as well as the
order of the phase transition which shows up within this framework).
Nevertheless we have assumed \mref{tens}
as a sensible phenomenological Ansatz
for the temperature dependence of the quark--quark interaction strength
and employed it in our model potential \mref{potenziale}
in order to describe the dynamical
evolution of the system at finite temperature.

In Section 2 we shall shortly review the many--body formalism employed to
deal with the system at zero temperature, introducing the relevant
quantities which might signal, as a function of the density, the occurrence
of the above mentioned phase transition. Numerical results are reported
at $T=0$ in order to better appreciate the modifications introduced by finite
values of the temperature. The latter are explored in Section 3, where a few
schematic details of finite temperature field theory precede the description
of the many--quark dynamics at finite temperature. The results for the ground
state energy per particle and for the pair correlation function are presented
at various (increasing) temperatures within the framework of the independent
pair approximation. Finally Section 4 summarizes the virtues and limitations
of the present approach.

\section{Results in the $T=0$ limit}
\setcounter{equation}{0}

Before introducing the finite temperature description, we shall report here
an outline of the results obtained at zero temperature. We have mainly
investigated the ground state energy (per particle) and the pair correlation
function at different values of the density.

The medium induced correlations have been taken into account by solving the
Bethe--Goldstone equation for the wavefunction of an interacting pair,
within the independent pair approximation. For this purpose we expand the
(antisymmetrized) two--particle wavefunction as follows:
%
\begin{eqnarray}
\Psi_{k_1k_2}(x_1,x_2)&\equiv&\varphi^{CM}_K(X)\psi_{nk}(x)\nonumber\\
& =&{\cal N}\frac{e^{iKX}}{\sqrt{L}}
\left\{c_{n0}\sin(kx)+\sum_{k_{1j},k_{2j}>F}c_{nj}\sin(k_{j}x)\right\},
\label{bgwf}
\end{eqnarray}
where $X=(x_1+x_2)/2$, $x=x_1-x_2$, $k_j=(k_{1j}-k_{2j})/2$ and the total
momentum $K=k_{1j}+k_{2j}$ is conserved in all terms in the expansion. The
restriction on $k_{1j}$ and $k_{2j}$ to be above the Fermi momentum $k_F$
accounts for the Pauli blocking of the states available to the considered
pair, which is due to the presence of the medium. In the present
one--dimensional calculation it can be rewritten as:
%
\begin{equation}
k_j > k_F+|K|/2.
\label{pauli}
\end{equation}

The trial wavefunction \mref{bgwf} is then inserted into the Schr\"odinger
equation for the relative motion with the potential \mref{potenziale}:
%
\begin{equation}
\left[-{d^2\over{dx^2}} + V(x)\right]\psi_{nk}(x)= E_{nk}\psi_{nk}(x),
\label{schro}
\end{equation}
which, by exploiting the orthogonality of the unperturbed wavefunctions
in \mref{bgwf}, reduces to a system of linear equations for the coefficients
$c_{nj}$ of the Bethe--Goldstone solution we are looking for. One should also
keep in mind that we are interested in bound (or quasi--bound) states of the
interacting pair, at least for those (low) densities where the confining
potential \mref{potenziale}
is strong enough to actually provide confinement. In order to
disentangle the few discrete eigenvalues of \mref{schro} out of its continuum
spectrum, we impose on $\psi_{nk}(x)$ the boundary condition
%
\begin{equation}
\psi_{nk}(R)=0,
\label{zeroinR}
\end{equation}
$R$ being an arbitrary length, of the order of (or larger than) the peak
position of the potential barrier developed by \mref{potenziale}
 at intermediate distances.
Eq. \mref{zeroinR} is easily satisfied by choosing the wavenumbers in the
expansion \mref{bgwf}
 to be integer multiples of $\pi/R$. If $\psi_{nk}$ corresponds to a
truly bound state, then its vanishing at $|x|=R$ will not be accidental and
we expect such a solution (and the corresponding energy eigenvalues) to be
fairly stable with respect to broad variations of R.

At zero temperature and very low densities $(\rho\le 0.1)$ we have
found \cite{abmn}
that the Bethe--Goldstone wavefunctions are close to the
bound states one can obtain in the pure Schr\"odinger equation (namely in the
absence of Pauli blocking), although they develop some small components
outside the ``confinement'' region. The importance of these components
rapidly increases with the density, thus showing that the effect of the
medium on the interacting pair loosens its binding and produces a gradual
transition (at intermediate densities) to a phase of non--interacting
particles (Fermi gas).

By summing over the energy eigenvalues of particles in occupied states,
namely those with relative momentum $k_\ell\le k_F-|K|/2$, one can evaluate
the ground state energy of the system: its evolution as a function of the
density is shown in Fig.~1, where it is compared with the purely kinetic
energy of a Fermi gas. The two coincide for large densities (say $\rho>1.0$)
but in the low density range the effect of correlations allows to disentangle
the existence of dynamically bound pairs, whose energy is fairly independent
upon the density of the system.

\begin{figure}[tb]
\vskip -2cm
\epsfxsize 11cm
\epsfysize 14cm
\centerline{\epsffile{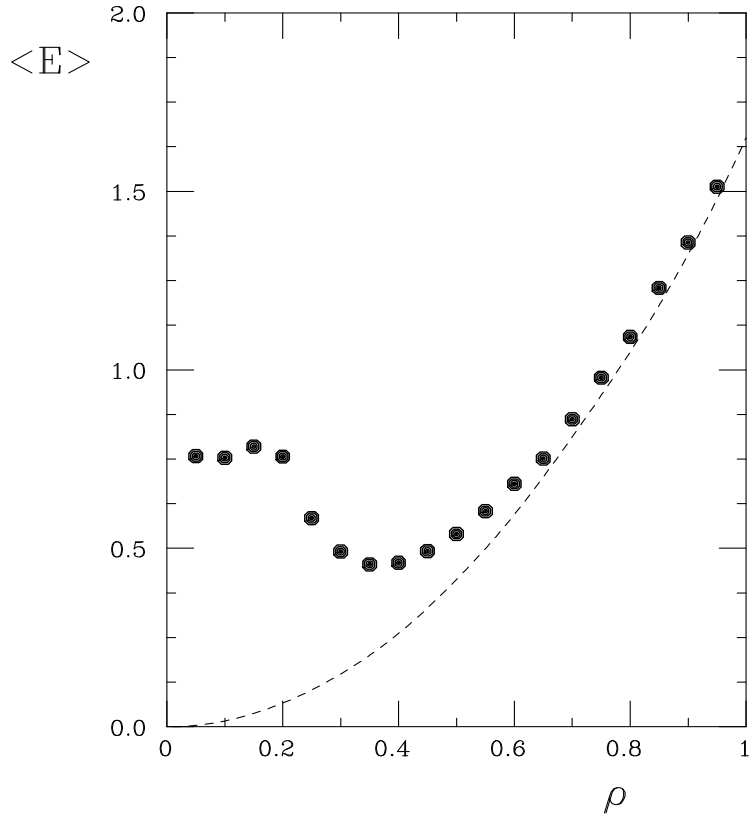}}
\vskip -2.2cm
\caption[Fig.1]{\parbox[t]{10cm}{\footnotesize
The ground state energy per particle at $T=0$ calculated
in the independent pair approximation (black dots) is shown as a function
of density, together with the one of a free Fermi gas (dashed line).}}
\end{figure}

Another quantity which provides a clear signature for the existence (if any)
of bound clusters in the system is the so--called pair correlation function
($r=|x|$)
%
\begin{equation}
g(r) =\frac{N(N-1)}{\rho^2}<\Psi|\rho_2(x_1-x_2)|\Psi>,
\label{gidierre}
\end{equation}
which is obtained as the ground state expectation value of the two--body
density operator:
%
\begin{equation}
\rho_2(x_1-x_2)=\frac{1}{N(N-1)}\sum_{i\ne j}\delta(x_i-x_1)
\delta(x_j-x_2).
\end{equation}
Within the independent pair approximation $g(r)$ turns out to be:
%
\begin{equation}
\gbg={1\over{\rho^2 L}}\sum_{|k_1|,|k_2|\le
k_F}|\psi_{0k}(x)|^2
\qquad\quad [k=(k_1-k_2)/2]
\label{bgcf}
\end{equation}
and can thus be evaluated using the ground state Bethe--Goldstone
wavefunctions \mref{bgwf}. In Fig.~2 we display a few $\gbg$ at $T=0$ and for
different densities: one can appreciate the evolution of the pair correlation
function from the low--density regime, where confinement is evident up to
$\rho\simeq 0.15$ (lower densities, not displayed here, show even higher
peaks at small distances), to the transition density ($\rho\simeq 0.19$),
where the Fermi gas component starts developing at large distances, up to
densities of the order of 0.5 and higher, where the pair correlation
function practically coincides with the one of an uncorrelated Fermi gas.
The correlation function obtained within the present model compares
fairly	well with the one of ref. \cite{Hor3},
 thus showing the equivalence
of the two approaches for what concerns the physical properties of
the system.

\begin{figure}[htb]
\vskip -2cm
\epsfxsize 11cm
\epsfysize 14cm
\centerline{\epsffile{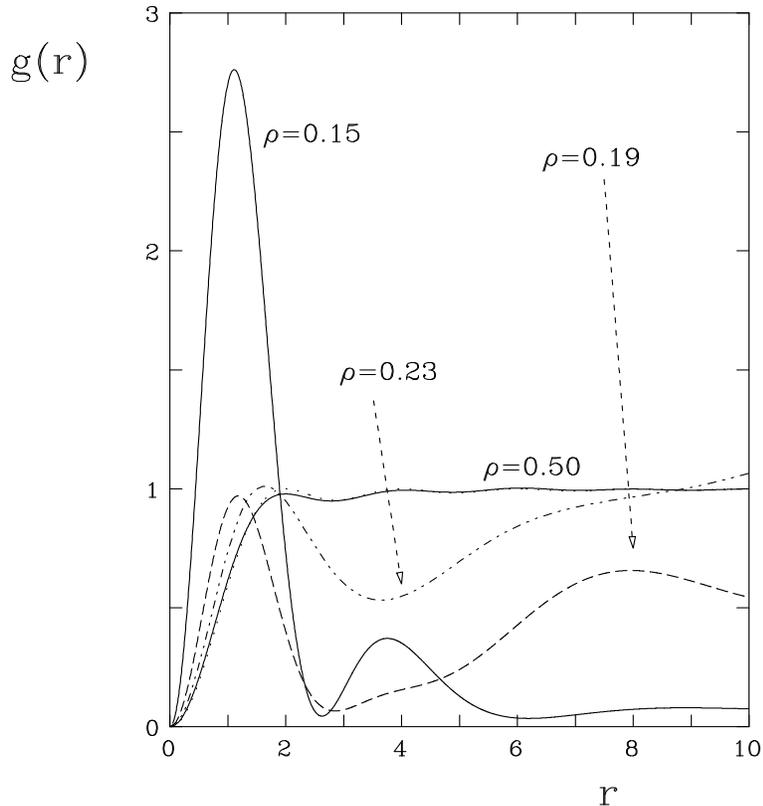}}
\vskip -2.2cm
\caption[Fig.2]{\parbox[t]{10cm}{\footnotesize
The pair correlation function $g(r)$ derived from the
Bethe--Goldstone wave functions at $T=0$ is displayed as a function of the
relative interquark distance: the various curves are labelled by the value
of the density. The free Fermi gas correlation function is shown only for
$\rho=0.5$ (dotted line), where it practically coincides with the one of the
correlated system.
}}\end{figure}

It is worth noticing that, in the presence of strong {\it short range}
interactions, the importance of the medium grows with the density (which
is proportional to $k_F$); in the present model, however, the interparticle
potential exponentially decreases with the density and thus the largest
modifications on the relative motion of a pair of particles only occur
within an intermediate, limited range of densities. This is an interesting
interplay between Pauli and dynamical correlations.

\section{Finite temperatures}
\setcounter{equation}{0}

In order to deal consistently with the hadron--quark plasma phase transition
one should take into account the behaviour of the system with the temperature,
thus extending the microscopic description within the formalism of finite
temperature field theory. In this case it is convenient to
statistically treat the system within the grand canonical ensemble, by
defining
%
\begin{equation}
K=H-\mu\hN,
\end{equation}
$\hN$ being the number operator and $\mu$ the chemical potential.

The expectation value of any operator will then be evaluated by implementing
the ensemble average
%
\begin{equation}
<O>={\rm Tr}(\hrho_G O),
\end{equation}
$\rho_G$ being the statistical density matrix $(\beta=1/k_B T)$
%
\begin{equation}
\hrho_G={1\over Z_G}e^{-\beta K}
\end{equation}
and $Z_G$ the grand partition function
%
\begin{equation}
Z_G\equiv e^{-\beta\Omega} = {\rm Tr} e^{-\beta K}.
\end{equation}
In the above the trace (Tr) implies a sum over a complete set of states
in the Hilbert space with any number of particles.

Moreover one can introduce the so--called {\it temperature} Green's
functions: for example the single particle propagator is defined as
%
\begin{equation}
\cG_{\alpha\beta}(\xb\tau,\xb'\tau')= -{\rm Tr}\left\{\hrho_G
T_\tau\left[{\hat\psi}_{K\alpha}(\xb\tau)
{\hat\psi}^\dagger_{K\beta}(\xb'\tau')\right]\right\},
\end{equation}
where ${\hat\psi}(\xb\tau)$ is the imaginary--time field operator in the
(modified) Heisenberg picture \cite{FetW}
 and $T_\tau$ the corresponding time
ordering operator.

For a non--interacting system of fermions the temperature Green's
function reads then (in momentum space):
%
\begin{equation}
\cG^0_{\alpha\beta}(\kb,\om;T)=\delta_{\alpha\beta}\left[
{n^0_k(T)\over{\om-\left(\eps^0_k-\mu\right)+i\eta}}
+ {{1-n^0_k(T)}\over{\om-\left(\eps^0_k-\mu\right)-i\eta}}\right]
\end{equation}
where the occupation probability
%
\begin{equation}
n^0_k(T)={1\over{1+e^{\beta(\eps^0_k-\mu)}}}
\label{fermi}
\end{equation}
reduces to the sharp theta function momentum distribution of the Fermi gas
in the limit of zero temperature:
%
\begin{equation}
n^0_k(T)\quad
\buildrel \longrightarrow \over {\scriptscriptstyle{T\rightarrow  0}}
\quad \theta(\eps^0_F-\eps^0_k).
\end{equation}

These ingredients can be used to evaluate, at $T\ne 0$, the polarization
(particle--hole) propagator within the customary perturbation theory and,
from it, to extract the two--body correlation function
according to the procedure described in ref. \cite{abmn}.

It should be pointed out that, in principle, the chemical potential has to
be self--consistently determined in order to satisfy the relation:
%
\begin{equation}
{N\over L}\equiv\rho={1\over{\sqrt{2}\pi}}\int_0^\infty {d\eps\over
{\sqrt{\eps}[1+\exp\{\beta(\eps-\mu)\}]}},
\label{chempot}
\end{equation}
which, in the limit of very small temperatures leads, for a free Fermi gas,
to the expansion:
%
\begin{equation}
\mu\simeq \eps_F\left[ 1+{1\over 12}\left({\pi\over{\beta\mu}}\right)
+\cdots\right].
\end{equation}
The latter, however, becomes unreliable for temperatures of the order of
the Fermi energy, which are typical values we are interested in; thus we
have utilized the chemical potential provided by equation \mref{chempot}.

Concerning the evaluation of the pair correlation function,
the alternative method we have employed here (in analogy with the previous
Section) consists in solving the Bethe--Goldstone equation for the
relative motion of two interacting quarks. At $T\ne 0$ the Pauli operator,
which limits the available states in the formal development \mref{bgwf} of the
correlated wavefunction, will be modified by employing the smooth
distribution function \mref{fermi}:
%
\begin{equation}
\theta(k_\alpha-k_F)\theta(k_\beta-k_F) \quad
\buildrel \Longrightarrow \over {\scriptscriptstyle{T \ne 0}}
\quad \left[1-n^0_{k_\alpha}(T)\right]\left[1-n^0_{k_\beta}(T)\right].
\label{fermibg}
\end{equation}
As a result the effect of Pauli correlations is somewhat weakened and
the influence of the medium can be expected to be, at a fixed density,
less important at finite temperature than at $T=0$.

\begin{figure}[htb]
\vskip -2cm
\epsfxsize 11cm
\epsfysize 14cm
\centerline{\epsffile{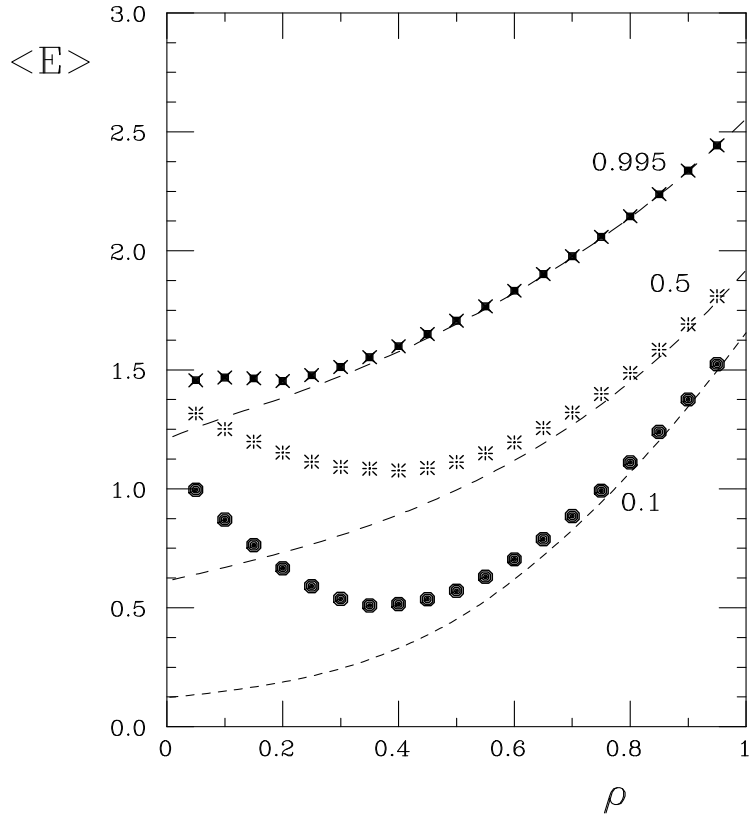}}
\vskip -2.2cm
\caption[Fig.3]{\parbox[t]{10cm}{\footnotesize
The ground state energy per particle at various (finite)
temperatures, calculated in the independent pair approximation is shown as a
function of the density: black dots
correspond to $T/T_C=0.1$, stars to $T/T_C=0.5$ and crosses to $T/T_C=0.995$.
The dashed lines represent the corresponding free kinetic energy.
}}\end{figure}

Moreover, as already anticipated in the Introduction, as the temperature
increases, one should also take into account some temperature dependence
of the effective coupling, which we have assumed in the form suggested by
\mref{tens}. This potential has been employed in the evaluation
of the two--body correlation function derived from the solution of the
Bethe--Goldstone equation, where, as stated above, we also account for the
temperature dependence of the occupation probability, according to
\mref{fermibg}.

In addition, from the corresponding energy eigenvalues we have evaluated the
ground state energy per particle as a function of the density, to be compared
with the pure kinetic energy of the Fermi gas, and followed its evolution
with increasing temperature (see Fig.~3).
At high densities the system is clearly a gas of
non--interacting quarks; at low densities, instead, for small (and zero)
temperatures the effect of pair correlations is clearly visible, but
tends to fade away, as it is obvious from the temperature dependence of
the quark--quark interaction, as the temperature approaches the critical
value (here arbitrarily fixed ``a priori'' to coincide with twice the Fermi
energy at $\rho=0.5$, $T_C=\pi^2/4$). Notably the discontinuity
in the derivative of the energy per particle, which is seen at $T=0$
and $\rho\simeq 0.2$, is smoothed out at $T\ne 0$.

Although, in the present treatment, we cannot identify a specific order
parameter, which would allow to consider the phase transition from a
thermodynamical point of view, we have interpreted as a transition density,
$\rho_C$, the one where the {\it medium induced} correlations vanish: this
quantity can be defined as the average value (with respect to the relative
and total momentum of a pair) of the difference between the matrix elements
of the bare potential \mref{potenziale} and the G-matrix which one obtains
from the solution of the Bethe--Goldstone equation:
%
\begin{equation}
\Delta U(\rho) ={\overline {<k,K|V|k,K>}} - {\overline {<k,K|G|k,K>}}.
\label{deltau}
\end{equation}
In the above, according to the usual definition,
%
\begin{equation}
<k,K|G|k,K>= {1\over R}\int_0^R dx \sin(kx)V(x)\psi_{0k}(x),
\label{gmat}
\end{equation}
where both the unperturbed and the correlated wavefunctions are normalized
over the above mentioned distance $R$ and the integral over the center of mass
coordinate is unity; the dependence upon $K$ of the right hand side of
\mref{gmat}
is implicit in the Bethe--Goldstone wavefunction through the action of the
Pauli operator.

The choice of $\Delta U$ as an ``order parameter'' is arbitrary,
but it is closely related to the energy gap usually considered in the
microscopic description of superconductivity; in that case a non--vanishing
gap signals the existence of bound electron pairs which profoundly alter
the global properties of the system. Here we assume \mref{deltau} as a
discriminant
between a system of ``hadrons'' (bound pairs of quarks) and a weakly
interacting Fermi gas of quarks. For each value of the temperature, the
solution of the equation $\Delta U(\rho)=0$ provides the critical density
for the deconfinement phase transition.

\begin{figure}[htb]
\vskip -2cm
\epsfxsize 11cm
\epsfysize 14cm
\centerline{\epsffile{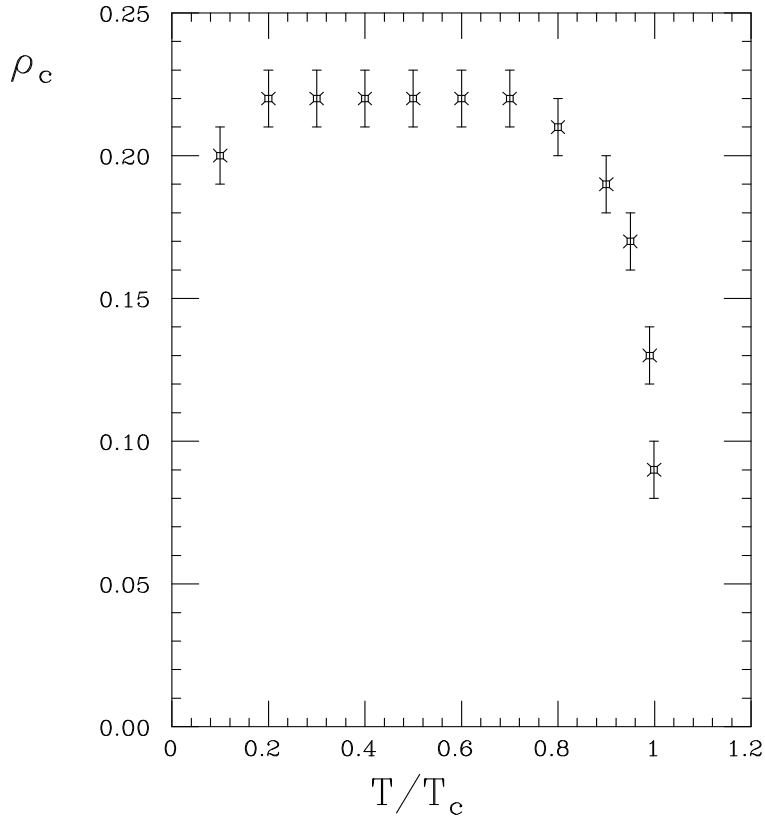}}
\vskip -2.2cm
\caption[Fig.4]{\parbox[t]{10cm}{\footnotesize
The transition density versus temperature (in units of $T_C$),
as it can be deduced by the quantity $\Delta U$: the error bars are inherent
to the averaging procedure employed in the determination of $\rho_C$.
}}\end{figure}

The behaviour of $\rho_C$ with the temperature is shown in Fig.~4, which
can be interpreted as the phase diagram for the model under investigation.
For a wide range of temperatures the critical density remains fairly
constant, then it rapidly drops to zero as one approaches the critical
temperature. It should be noticed the slight increase of $\rho_C$ in the
low temperature regime: this outcome, somewhat opposite to the intuitive
expectation, stems from the fact that a non--vanishing temperature
weakens the Pauli blocking on the correlated wavefunction while the
interaction strength is still practically unaffected. Thus for low
temperatures the dominant effect is the appearance of the Fermi gas
component at densities higher than for $T=0$; as the temperature increases,
however, this tendency is balanced and overcome by the temperature
dependence of the interaction itself.

\begin{figure}[htb]
\vskip -2cm
\epsfxsize 11cm
\epsfysize 14cm
\centerline{\epsffile{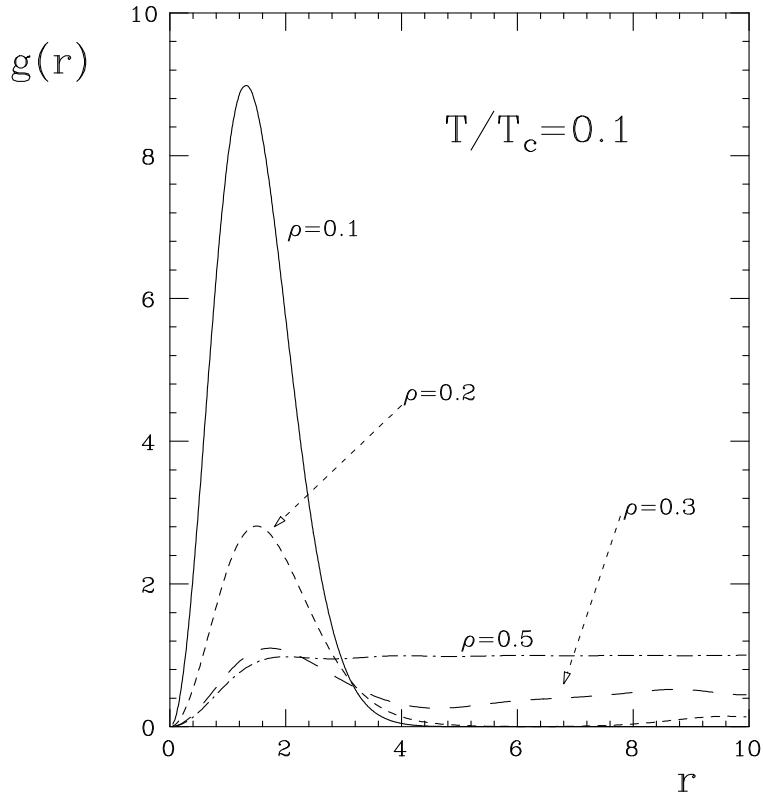}}
\vskip -2.2cm
\caption[Fig.5]{\parbox[t]{10cm}{\footnotesize
The pair correlation function $g(r)$ derived from the
Bethe--Goldstone wave functions at $T/T_C=0.1$ is displayed as a function of
the relative interquark distance: the various curves are labelled by the value
of the density.
}}\end{figure}
\begin{figure}[htb]
\vskip -2cm
\epsfxsize 11cm
\epsfysize 14cm
\centerline{\epsffile{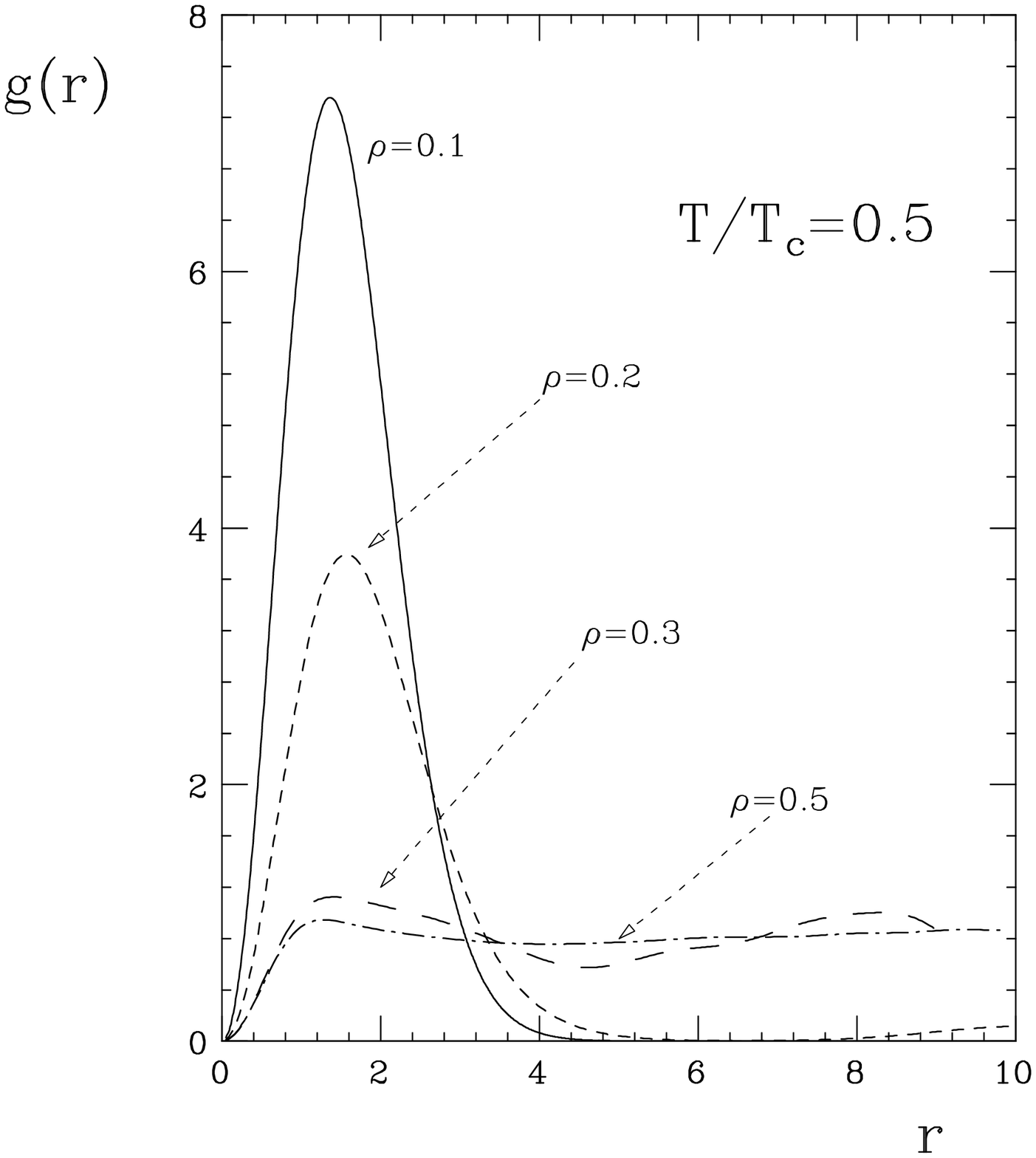}}
\vskip -2cm
\caption[Fig.6]{\footnotesize The same as
in Fig.~5, at $T/T_C=0.5$}\end{figure}
\begin{figure}[htb]
\vskip -2.2cm
\epsfxsize 11cm
\epsfysize 14cm
\centerline{\epsffile{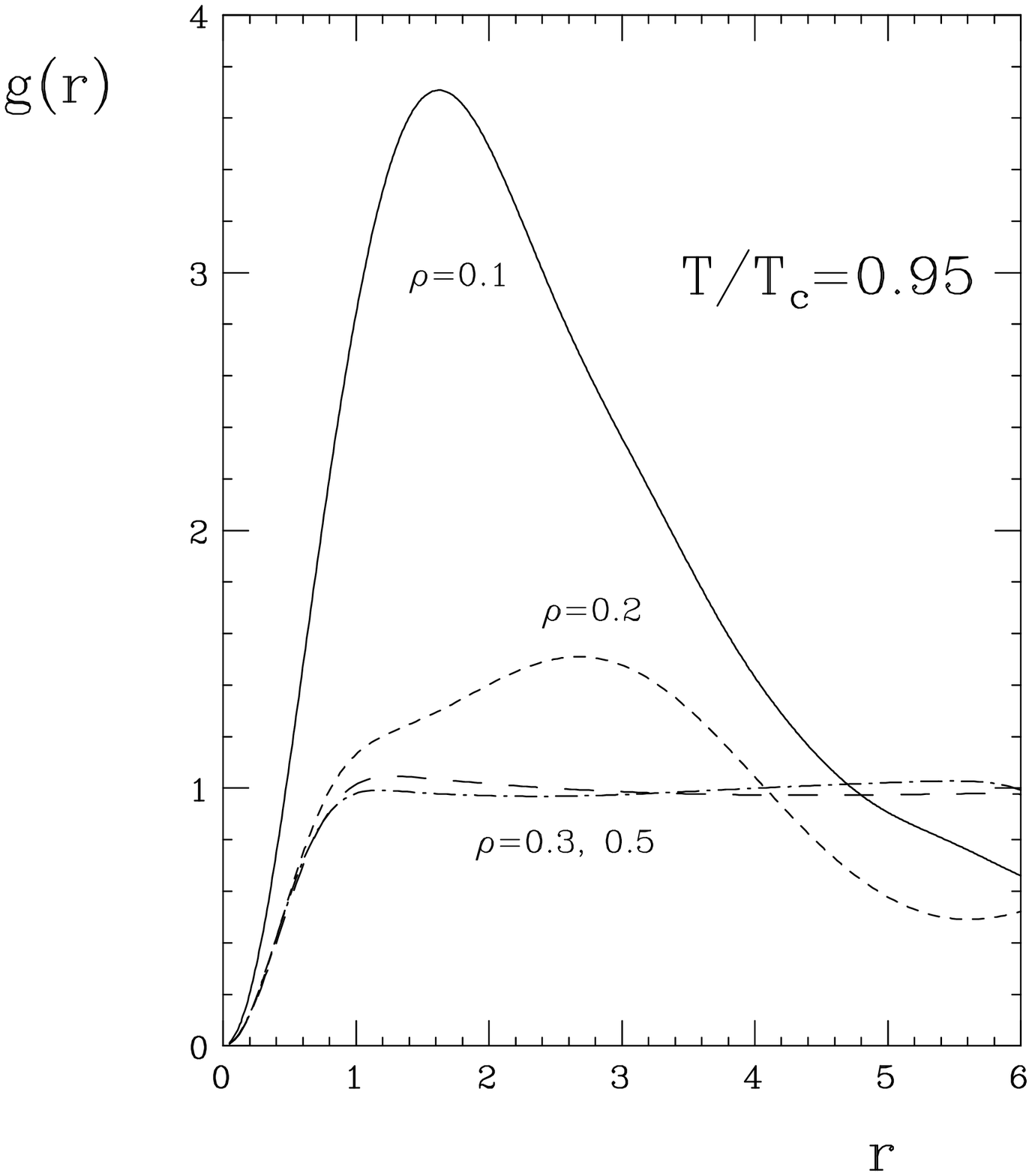}}
\vskip -2.2cm
\caption[Fig.7]{\footnotesize
The same as in Fig.~5, at $T/T_C=0.95$}\end{figure}

Finally we have evaluated the pair correlation function, eq.~\mref{bgcf}, by
utilizing the finite temperature Bethe--Goldstone wavefunctions for the
relative motion. A few examples of $g(r)$ at various densities are
illustrated in Figs.~5--7 for typical values of the temperature (in units
of the above defined critical temperature). For low $T$ ($T=0.1T_C$) the
same behaviour as for $T=0$ is observed, although, in agreement with the
phase diagram of Fig.~4, the transition density appears to be somewhat
larger than in the zero temperature limit. As the temperature increases
the binding effects become weaker and weaker and the correlation function
rapidly approaches the free Fermi gas one, since the quark--quark model
interaction becomes vanishingly small. Notice that the finite temperature
washes out completely the quantum oscillations of the free correlation
function.

\section{Remarks and conclusions}

In this work we have investigated some specific properties of a system
of fermions, strongly interacting with a density dependent force, such
as to provide bound pairs in the low density limit e a free Fermi gas
at large densities. As already pointed out in a previous work \cite{abmn},
this could bear some relevance in the investigation of the evolution of
the primordial quark--gluon plasma toward the hadronic phase. However
the evolution of the early Universe non only implies a decreasing local
barionic density but also a rapid fall off of the temperature: the latter
indeed should play a non--negligible role in the phase transition which
lead to the formation of hadrons and nuclei.

Here we have thus focussed the attention on the effects of temperature,
starting from the hypothesis (supported by lattice QCD calculations)
that with increasing temperature the strength of the confining potential
becomes weaker and finally vanishes at some critical temperature. On this
basis we have analyzed the interplay of density and temperature dependences
of the quark--quark interaction: the pair correlation function shows indeed
that, with increasing temperature, the transition density remains
fairly stable until the critical temperature is approached and then
rapidly drops to zero.

The above findings show that the present approach is on the right path
to achieve a microscopic description of the phase transition from hadrons
to a quark--(gluon) plasma. However it can only be considered as a very
preliminary stage: indeed the model for quark matter employed here
suffers from many major limitations. Beside being one--dimensional and
non--relativistic, it does not embody internal degrees of freedom like
spin, colour, etc. Also it does not contain explicit gluonic degrees of
freedom, not even in the form of a perturbative one gluon exchange.

Moreover, with respect to the present and future experiments of relativistic
heavy ion collisions, one would also like to describe the dynamical evolution
with time of the many--quark system, so that estimates of the probability
for QGP formation and subsequent hadronization could be done.
Future work is required in order to improve along these directions.


\vfill

\begin{thebibliography}{99}
\bibitem{Hor1} C.J. Horowitz, E.J. Moniz and J.W. Negele, Phys. Rev.
{\bf D 31}, 1689 (1985);
\bibitem{Hor2} C.J. Horowitz and J. Piekarewicz, Nucl. Phys. {\bf A 536}, 669
(1992);
\bibitem{Hor3} C.J. Horowitz and J. Piekarewicz, Phys. Rev. {\bf C 44}, 2753
(1991);
\bibitem{abmp} W.M. Alberico, M.B. Barbaro, A. Molinari and F. Palumbo, Z.
Phys.
{\bf A 341}, 327 (1992);
\bibitem{Fried} R. Friedberg and T.D. Lee, Phys. Rev. {\bf D 15}, 1694;
{\bf D 16}, 1096 (1977);
\bibitem{Li} Ming Li, M.C. Birse and L. Wilets, J. Phys. G: Nucl. Phys.
{\bf 13}, 1 (1987);
\bibitem{Chan} G. Chanfray, O. Nachtmann and H.J. Pirner, Phys. Lett.
{\bf B147}, 249 (1984);
\bibitem{Pir} H.J. Pirner, Prog. Part. Nucl. Phys. {\bf 29}, 33 (1992);
\bibitem{Pie} J. Piekarewicz and J.R. Shepard, Phys. Rev. {\bf C 45}, 2963
(1992);
\bibitem{Pis} R.D. Pisarski and O.Alvarez, Phys. Rev. {\bf D 26}, 3735 (1982);
\bibitem{abmn} W.M. Alberico, M.B. Barbaro, A. Magni and M. Nardi, Nucl. Phys.
{\bf A 552}, 495 (1993);
\bibitem{FetW} see, for example, A.L. Fetter and J.D. Walecka, {\it Quantum
Theory of Many Particle Systems} (Mc Graw--Hill Book Co., 1971), Ch.7;
\end{thebibliography}
\end{document}